\documentclass[twocolumn]{article}
\usepackage{natbib}

\title{Intrinsic Origin Of Extreme-Scale Rotation Of Quasar
Polarization Vectors}

\author{N.A. Silant'ev, M.Yu. Piotrovich,
 Yu.N. Gnedin\thanks{E-mail: gnedin@gao.spb.ru},
 T.M. Natsvlishvili\\ Central Astronomical Observatory at
 Pulkovo, 196140, Saint-Petersburg, Russia.}

\begin{document}

\maketitle

\begin{abstract}
Extreme-scale alignment of  quasar optical polarization vectors at
cosmological scales ($z\le 2$) is also characterized by the
rotation of mean position angle $\chi$ with $\Delta \chi \approx
30^{\circ}$ per 1 Gpc. For observing interval of $z$ the total
rotation angle acquires the value $\sim 90^{\circ}$. We suggest
the possible explanation of the half of this rotation as a
consequence of physical transformation of initially vertical
magnetic field ${\bf B}_{\|}$, directed along the normal ${\bf N}$
to the surface of accretion disk, into the horizontal
(perpendicular to ${\bf N}$) one. We found asymptotical analytical
expressions for axially averaged polarization degree $p$ and mean
position angle $\chi$ for various types of magnetized accretion
disks. We found also that during the evolution can be realized the
case $B_{\bot}\approx B_{\|}$ where position angle $\chi$ rotates
from $45^{\circ}$ to zero. This rotation may occur during fairly
great cosmological time (corresponding to $\Delta z\sim 1-2$). The
part of rotation $\sim \Delta \chi \approx 45^{\circ}$ can be
explained by a mechanism of alignment of polarization vectors, say
distribution of the part of quasars as a spiral in the cosmic
space with slow variation of rotation axis of corresponding
accretion disks. Both mechanisms are mutually related one with
another.

{\bf Keywords:} polarization, magnetic fields, accretion disks,
quasars, active galactic nuclei.
\end{abstract}

\section{Introduction}

Large scale alignments of quasar polarization vectors have been
revealed by \citet{hutsemekers98}, looking at a sample of 170 QSO
selected from various surveys. \citet{hutsemekers01} have
confirmed this effect later on a larger sample. The departure from
random orientation has been found at fairly well significance
level. Hutsemekers and Lamy have concluded that these alignments
seemed to come from high redshift regions, implying that the
underlying mechanism might cover physical distances of
gigaparsecs. Moreover, \citet{hutsemekers05} have estimated the
rotation of mean position angle magnitude as $\Delta \chi \leq
30^{\circ}$ at the distance of $\sim 1$Gpc. In this paper they
presented the analysis of the alignment effect for a total sample
of 355 quasars, comprising new polarization measurements both from
observing runs 2001-2003 and new comprehensive data from surveys
and the literature. The paper of \citet{borquet08} threw light on
the new observational fact. They found a significant correlation
between the polarization position angle and the position angle of
the major axis of the host galaxy extended emission.

The first question arises concerning contamination of interstellar
dust grains in our Galaxy. The linear dichroism of aligned
interstellar dust grains in our Galaxy produces really linear
polarization along the line of sight. This polarization
contaminates to some extent the quasar measured data and may
change their position angles. However, \citet{sluse05} have shown
that interstellar polarization has a little effect on the
polarization angle distribution of significantly polarized ($p \ge
0.6\%$) quasars.

The first explanation of observed large-scale alignment of quasar
polarization vectors was naturally associated to light propagation
into intergalactic medium. \citet{hutsemekers05} claimed that,
while interpretations like a global rotation of the Universe can
potentially explain this effect, the properties of observed
phenomenon correspond, at first glance, to the dichroism and
birefringence predicted, for example, by photon-pseudoscalar
(axion-like particle) oscillation within a intergalactic magnetic
field. This process has been considered in detail in a number of
papers \citep{jain03, das05, hutsemekers05, gnedin07,
piotrovich08, hutsemekers08, agarwal09}. The probability of
magnetic conversion of photons into low-mass pseudoscalar
particles was calculated in the papers \citep{anselm82, raffelt88,
harari92, gnedin92, raffelt96, deffayet01, csaki02, das05}.

However the basic difficulty arises because of uncertainty of
photon-pseudoscalar (axion) mixing constant. Unfortunately, the
famous CAST experiment gives only constraints on this mixing
\citep{andriamonje07}. Recently \citet{piotrovich08} have
estimated the photon-pseudoscalar mixing constant from the effect
of large-scale alignment and rotation of polarization plane of
distant ($z > 0.5$) quasars. Many authors have considered process
for generating magnetic fields of the cosmological interest. Their
conclusions allow us to consider unclear the magnetic field
strength interpretation of this results, in particular, whether
the rotation of polarization angle is associated with
intergalactic medium, or with the presence of some intrinsic
physical phenomena in the nearest environments of the polarized
quasars.

Here we present another explanation of the effect of extreme-scale
rotation position angle versus cosmological redshift $z$. We use
the results of calculation of polarization of radiation from
optically thick magnetized accretion disks in quasars and active
galactic nuclei. The presence of intrinsic magnetic field in an
accretion disk produces a new effect, provided by the Faraday
rotation of polarization plane along a photon mean free path in
scattering medium (see, for example, \citet{dolginov95};
\citet{gnedin97}; \citet{silantev02}; \citet{gnedin06}).

Due to these authors, a nontrivial wavelength dependence of
polarization and position angle arises when the Faraday rotation
angle $\Psi$ at the Thomson optical depth $\tau$ is sufficiently
large:

\begin{equation}
 \Psi = 0.4 \left(\frac{\lambda}{1\mu m}\right)^2
 \left(\frac{B}{1G}\right) \tau \cos{\theta}\equiv
 \frac{1}{2}\delta\tau\cos{\theta}.
 \label{eq1}
\end{equation}

\noindent Here, $\lambda$ is the radiation wavelength and $\theta$
is the angle between the line of sight ${\bf n}$ and magnetic
field ${\bf B}$. Below we shall use the known Milne problem which
corresponds to the case where the thermal sources are located far
from the surface of optically thick accretion disk. The existence
of Faraday rotation angle $\Psi$ gives rise to both the decrease
of degree of polarization $p$ and the rotation of position angle
$\chi$ (up to maximum value $\chi=45^{\circ}$ as compared the
usual Thomson scattering problem where $\chi =0$ is assumed). The
numerical solution of the Milne problem in the case of magnetized
atmosphere with magnetic field directed along the normal to
atmosphere has been early obtained by \citet{silantev94,
silantev02, agol96} and \citet{shternin03}.

Magnetic field in an accretion disk consists of two mutually
perpendicular components ${\bf B}={\bf B}_{z}+{\bf B}_{\bot}$.
Here ${\bf B}_z\equiv {\bf B}_{\|}$ is directed along the normal
${\bf N}$ to the disk's surface. The field ${\bf B}_{\bot}={\bf
B}_{\rho} +{\bf B}_{\varphi}$ is perpendicular to ${\bf N}$, and
consists of the azimuthal ${\bf B}_{\varphi}$ and radial (in
cylindrical system of references) ${\bf B}_{\rho}$ components.

We observe the azimuthally averaged Stokes parameters $\langle
Q({\bf n},{\bf B})\rangle$ and $\langle U({\bf n},{\bf
B})\rangle$. If ${\bf B}_{\|}=0$, the observed position angle
$\chi\equiv 0$ due to symmetry of a problem (remember that
$\tan{2\chi}=U/Q$, and the direction with $\chi=0$ is parallel to
disk's surface). In contrary, the case ${\bf B}_{\bot}=0$ gives
rise to the rotation of position angle $\chi$ from zero at
$B_{\|}=0$ up to maximum value $\chi_{max}=\pm 45^{\circ}$ for
large values of $B_{\|}$. The signs $\pm$ depend on outside or
inside the accretion disk orientation of magnetic field ${\bf
B}_{\|}$. In both cases the increase of magnetic field value
decreases the polarization degree $p$.

Note that Faraday rotation of polarization plane in intergalactic
medium is too low to be considered as a possible mechanism of
observed rotation of mean position angle.

The goal of the paper is to show that in general cases there exist
situations where the position angle $\chi$ decreases, with the
increase of ${\bf B}_{\bot}$, from initial value $\chi \approx
45^{\circ}$ (corresponding to large $B_{\|}$) up to $\chi \approx
0$. So, we shall demonstrate that the evolution of components
${\bf B}_{\|}$ and ${\bf B}_{\bot}$ can explain the observed
large-scale position angle rotation due to intrinsic physical
evolution of magnetic fields in accretion disks around of QSOs or
AGNs. Below we present this in detail. We shortly discuss also the
origin of possible mechanisms of transformation ${\bf B}_{\|}$ to
${\bf B}_{\bot}$.

\section{Basic equations}

Silant'ev (2002) obtained the analytical approximate formulae for
the Stokes parameters of polarization of radiation emerging from
optically thick accretion disk which for the Milne problem acquire
the form:

\[
 I = \frac{F}{2 \pi J_1} J(\mu),
\]
\[
 Q = - \frac{F}{2 \pi J_1} \frac{1 - g}{1 + g}
 \frac{(1 - \mu^2)(1 +C - k \mu)}{(1 +C - k \mu)^2 +
 (1-q)^2\delta^2 \cos^2\theta},
\]
\begin{equation}
 U = - \frac{F}{2 \pi J_1} \frac{1 - g}{1 + g}
 \frac{(1 - \mu^2)(1 - q) \delta \cos{\theta}}{(1 +C - k \mu)^2 +
 (1-q)^2\delta^2 \cos^2\theta}.
 \label{eq2}
\end{equation}

\noindent where $\theta$ is the angle between the directions of
magnetic field ${\bf B}$ and the line of sight ${\bf n}$, $\mu =
\cos{i}$ ($i$ is inclination angle, i.e. angle between normal
${\bf N}$ and ${\bf n}$), $q$ is the degree of true absorption: $q
=\sigma_a / (\sigma_a + \sigma_s)$. The function $J(\mu)$
describes the angular distribution of emerging radiation, $F$ is
the radiation flux emerging from the disk's surface. The function
$J(\mu)$ and the numerical parameters $g$, $J_1$, and $k$ are
tabulated by \citet{silantev02}.  The minus signs denote that
Thomson polarization is perpendicular to the plane $({\bf n}{\bf
N})$, i.e. has position angle $\chi=0$. In the case of dominant
electron scattering inside an accretion disk: $g = 0.83255,
J_1=1.19400$, and $k=0$. Dimensionless parameter $C$ arises in
turbulent magnetized plasma (see \citet{silantev05}) and
characterizes the new effect - additional extinction of parameters
$Q$ and $U$ due to incoherent Faraday rotations in small turbulent
eddies.

The parameters of the Faraday depolarization for $B_{\|}$ and
$B_{\bot}$ can be introduced from Eq.(\ref{eq1}):

\[
 \delta_{\|} = 0.8 \left(\frac{\lambda_{rest}}{1 \mu m}\right)^2
 \left(\frac{B_{\|}}{1 G}\right),
\]
\begin{equation}
 \delta_{\bot} = 0.8 \left(\frac{\lambda_{rest}}{1 \mu m}\right)^2
 \left(\frac{B_{\bot}}{1 G}\right).
 \label{eq3}
\end{equation}

\noindent The wavelength $\lambda_{rest}$ is derived in the rest
system of quasar or AGN. The value $\delta\cos\theta$ in these
notions has very simple form:

\begin{equation}
 \delta\cos\theta=a+b\cos\Phi,
 \label{eq4}
\end{equation}

\noindent where dimensionless parameters $a$ and $b$ are connected
with the parameters $\delta_{\|}$ and $\delta_{\bot}$

\begin{equation}
 a =(1-q) \delta_{\|} \mu, \,\,\,\,\,b=(1-q)\delta_{\bot}
 \sqrt{1-\mu^2},
 \label{eq5}
\end{equation}

\noindent with the angle $\Phi=\phi +\phi_*$, where $\phi$ being
the azimuthal angle of radius-vector of observed point ${\bf r}$
on the surface of an accretion disk. The angle $\phi_*$ is the
angle between ${\bf B}_{\bot}$ and ${\bf B}_{\rho}$. The azimuthal
angle of line of sight ${\bf n}$ is taken zero.

Non-polarized light escape the optically thick disk basically from
the surface layer with $\tau \approx 1$.  The additional
extinction of parameters $Q$ and $U$ in turbulent magnetized
atmosphere means that linearly polarized light escape the disk
surface mainly from the Thomson optical depth $\tau \approx
1/(1+C)$. If the Faraday rotation angle $\Psi$ corresponding to
this optical length becomes greater than unity, then the emerging
radiation will be depolarized as a result of summarizing of
radiation fluxes with very different angles of Faraday rotation.
For directions that are near perpendicular to the direction of the
magnetic field in an accretion disk the Faraday rotation angle is
too small to yield depolarization effect. Certainly, the diffusion
of radiation in the inner parts of a disk depolarizes it even in
the absence of magnetic field because of multiple scattering of
photons. The Faraday rotation only increases the depolarization
process. It means that the polarization of outgoing radiation
acquires the peak-like angular dependence with its maximum for the
direction perpendicular to the magnetic field. The sharpness of
the peak increases with increasing magnetic field strength. The
basic region of allowed angles appears to be $\sim 1 / \delta$.
Another very important feature characterizing the polarized
radiation is the wavelength dependence of polarization degree that
is strongly different from that for Thomson scattering.

Formulae (\ref{eq2}) give rise to the following expression for the
polarization degree:

\begin{equation}
 p(\mu, {\bf B}) =
 \frac{p(\mu, 0)}{\sqrt{(1+C-k\mu)^2 + (1 - q)^2 \delta^2
 \cos^2{\theta}}},
 \label{eq6}
\end{equation}

\noindent where $p(\mu,0)$ means the polarization degree for pure
Thomson scattering. Remind that $p(\mu, 0)$ in conservative
atmosphere (q=0) has the classical Sobolev-Chandrasekhar value,
that is maximal and equal to $11.7\%$ for $\mu = 0$ (i.e. for the
inclination angle $i = 90^{\circ}$). We use the reference system
with X-axis lying in the plane $({\bf nN})$. In this system
$U(\mu,0)=0$ and $p(\mu,0)=|Q(\mu,0)|$. The position angle $\chi$
of emerging radiation, as usually, is described by the known
relation:

\begin{equation}
 \tan{2\chi}=\frac{U}{Q}=\frac{(1-q)\delta\cos\theta}{1+C-k\mu}.
 \label{eq7}
\end{equation}

For strong magnetic field strength (or large wavelength,) when
$(1-q)\delta \cos{\theta} \gg 1$, the simple asymptotic
expressions take place:

\begin{equation}
 p(\mu, {\bf B}) \approx \frac{p(\mu, 0)}{(1-q)\delta
 \cos{\theta}},\,\,\,\,\, \chi\to 45^{\circ}.
 \label{eq8}
\end{equation}

It is seen from Eq.(\ref{eq6}) that small-scale magnetic
turbulence (parameter $C$) decreases the observed polarization
degree. This is because the polarized light (parameters $Q$ and
$U$) escape mainly from the level $\tau\approx 1/(1+C)$ where
intensity is lesser as compared with the level $\tau\approx 1$,
corresponding to escape of non-polarized light. In contrast, the
existence of absorption ($q\neq 0, k\neq 0$) increases
polarization both due to existence the parameter $k$ in the
denominator of Eq.~(6) and due to that $p(\mu, q,0)$ is higher
than $p(\mu,q=0, 0)$ (see, for example, Silant'ev (1980)). The
absorption gives rise to more sharp intensity (along the normal
${\bf N}$), and the situation look like the single scattering of a
light beam in the surface layer of the atmosphere. So, at $q=0.1$
the Milne problem has $J(\mu=1)=4.39$ and $p(\mu=0,0)=20.4\%$. The
corresponding values for conservative atmosphere are 3.06 and
11.71\%.

\section{Polarization degree and position angle of observed
accretion disk radiation}

For an accretion disk the light depolarization depends on the
geometry of magnetic field. Usually one observes the axially
symmetric accretion disks as whole. In this case the observed
integral Stokes parameters $\langle Q({\bf n},{\bf B}) \rangle$
and $\langle U({\bf n},{\bf B}) \rangle$ are described by the
azimuthal averaged expressions. To obtain analytical formulae for
$\langle Q\rangle$ and $\langle U\rangle$ we present  expressions
(2) for $Q$ and $U$ in a complex form:

\begin{equation}
 Q-iU=\frac{p(\mu,0)}{G +ia+ib\cos\Phi}.
 \label{eq9}
\end{equation}

\noindent Here and what follow we use the notion $G=1+C-k\mu$. The
azimuthally averaging of this formula gives rise to expression:

\begin{equation}
 \langle Q\rangle -i\langle U\rangle=\frac{p(\mu,0)}{2\pi(G+ia)}
 \int_0^{2\pi}d\Phi\frac{1} {(1+\epsilon\cos\Phi)}.
 \label{eq10}
\end{equation}

\noindent Here $\epsilon=ib/(G + ia)$. If parameter $b\neq 0$ the
$\Phi$-integration can be easily evaluated by residue theorem
(see, for example, \citet{smirnov64}). As a result, we obtain the
following expression:

\begin{equation}
 \langle Q\rangle -i\langle U\rangle =\pm \frac{p(\mu,0)}
 {\sqrt{G^2+b^2-a^2 +2iGa}},
 \label{eq11}
\end{equation}

\noindent where  sign plus corresponds to $|\epsilon|<1$, and
minus corresponds to $|\epsilon|>1$.

\citet{silantev05} has derived the next expression for the
turbulent extinction parameter:

\begin{equation}
 C = 0.64(1-q) \tau_1 \lambda_{rest}^4 (\mu m)\langle B'^{2}
 \rangle f_B / 3
 \label{eq12}
\end{equation}

\noindent Here, $\tau_1\ll 1$ is the mean Thomson optical length
of small turbulent eddies, the value $B'$ denotes the fluctuating
component of the magnetic field (${\bf B} = {\bf B}_0 + {\bf
B}'$). We omitted, for brevity, the subscript "0"in previous
formulae. It means that the depolarization parameters $a$ and $b$
in Eq.(\ref{eq4}) are determined only by the global magnetic field
values. The numerical coefficient $f_B \approx 1$ is connected
with the correlation function of fluctuating components ${\bf B'}$
in neighboring points of turbulent atmosphere.

It is convenient to introduce the relative polarization degree
$p_{rel}=p({\bf n},{\bf B})/p(\mu,0)$. By the usual way we obtain
from Eq.(\ref{eq11}) the following expressions:

\begin{equation}
 p_{rel}=\frac{1}{[G^4+2G^2(a^2+b^2)+(a^2-b^2)^2]^{1/4}}.
 \label{eq13}
\end{equation}

\noindent The relative polarization degree $p_{rel}$ does not
depend on signs $\pm$ in Eq.(\ref{eq11}). The interesting property
of expression (\ref{eq13}) that it depends symmetrically on
parameters $a$ and $b$, i.e. we have $p_{rel}(\mu,
a,b)=p_{rel}(\mu, b,a)$.

The position angle $\chi$ is connected with the phase angle $\phi$
of the complex expression inside the root in Eq.(\ref{eq11}).
Evidently we have

\begin{equation}
 \tan\phi=\frac{2Ga}{G^2+b^2-a^2}.
 \label{eq14}
\end{equation}

\noindent If right side in Eq.(\ref{eq14}) is positive (this
corresponds to sign plus in Eq.(\ref{eq11})), then the position
angle $\chi$ is determined by expression

\[
 \tan{2\chi}=\frac{\langle U\rangle}{\langle Q\rangle}=
\]
\begin{equation}
 \frac{2Ga}{\sqrt{G^4+2G^2(a^2+b^2)+(a^2-b^2)^2}+
 (G^2+b^2-a^2)}.
 \label{eq15}
\end{equation}

\noindent For the negative $2Ga/(G^2+b^2-a^2)<0$ we have

\[
 \tan{2\chi} =
\]
\begin{equation}
 = \frac{2Ga}{\sqrt{G^4+2G^2(a^2+b^2)+(a^2-b^2)^2}-
 |G^2+b^2-a^2|}.
 \label{eq16}
\end{equation}

\noindent Evidently the position angle $\chi$ is not symmetric
function of parameters $a$ and $b$. The observed parameters
$\langle U\rangle$ and $\langle Q\rangle$ are positive for our
choice of the reference frame and corresponds to the case where
${\bf B}$ is directed outside the accretion disk surface. If the
magnetic field ${\bf B}$ is directed inside the accretion disk,
then the position angle $\chi $ changes its sign (parameter
$U<0$).

For a number of particular cases one can obtain fairly simple
analytical expressions. First of all, for the case of pure
vertical magnetic field (parameter $b=0$) we obtain directly from
basic Eq.(\ref{eq9}) the following expression:

\begin{equation}
 p_{rel}(\mu,{\bf B}_{\|}) = \frac{1}{\sqrt{G^2 + a^2}},\,\,\,\,
 \tan{2\chi} = \frac{a}{G}
 \label{eq17}
\end{equation}

\noindent For pure perpendicular magnetic field ($a=0$) our
general formulae give:

\begin{equation}
 p_{rel}(\mu,{\bf B}_{\bot})=\frac{1}{\sqrt{G^2+b^2}},\,\,\,\,
 \chi\equiv 0.
 \label{eq18}
\end{equation}

\noindent The position angle $\chi =0$ is due to symmetry of
problem in this case. For the case $a=b$, which approximately
corresponds to equipartition $B_{\|}=B_{\bot}$ for $i\approx
45^{\circ}$, the formulae for $p$ and $\chi$ acquire the forms:

\[
 p_{rel}(\mu, a=b)=\frac{1}{[G^2(G^2+4a^2)]^{1/4}}
\]
\begin{equation}
 \tan{2\chi}=\frac{2Ga}{\sqrt{G^2(G^2+4a^2)}+G^2}.
 \label{eq19}
\end{equation}

\noindent Remember that $G=1+C-k\mu$. It is easy check that for
conservative atmosphere ($k=0$) the existence of $b=a$ increases
the relative polarization degree $p_{rel}$ as compared with the
case pure parallel magnetic field ($a\neq 0, b=0$):

\begin{equation}
 \frac{p_{rel}(\mu, a,b=0)}{p_{rel}(\mu,a=b)}=
 \left(\frac{G^2+a^2}{G^2(G^2+4a^2)}\right)^{1/4}<1.
 \label{eq20}
\end{equation}

Below we present Tables 1 - 4, where $p_{rel}$ and $\chi $ are
given for different values of $a,b$ and $C$ in conservative
atmosphere ($q=0$).

\begin{table}
 \caption{The relative polarization degree
 $p_{rel}=p({\bf n},{\bf B})/p(\mu,0)$ as a function of
 parameters $a$ and $b$ at $C=0$.}
 \label{tab1}
 \begin{tabular}{@{}lccccccc}
 \hline
 $a\backslash b$ & 0     & 2     & 4     & 6     & 8     & 10    & 20 \\
 \hline
 0               & 1     & 0.447 & 0.242 & 0.164 & 0.124 & 0.100 & 0.050 \\
 2               & 0.447 & 0.492 & 0.271 & 0.173 & 0.128 & 0.101 & 0.050 \\
 4               & 0.242 & 0.271 & 0.352 & 0.211 & 0.142 & 0.108 & 0.050 \\
 6               & 0.164 & 0.173 & 0.211 & 0.288 & 0.178 & 0.123 & 0.051 \\
 8               & 0.124 & 0.128 & 0.142 & 0.178 & 0.250 & 0.158 & 0.054 \\
 10              & 0.100 & 0.102 & 0.108 & 0.123 & 0.158 & 0.223 & 0.058 \\
 20              & 0.050 & 0.050 & 0.051 & 0.052 & 0.054 & 0.058 & 0.149 \\
 \hline
\end{tabular}
\end{table}

\begin{table}
 \caption{The position angles $\chi^{\circ}$ as a
 function of parameters $a$ and $b$ at $C=0$.}
 \label{tab2}
 \begin{tabular}{@{}lccccccc}
 \hline
 $a\backslash b$ & 0     & 2    & 4    & 6    & 8    & 10   & 20 \\
 \hline
 0               & 0     & 0    & 0    & 0    & 0    & 0    & 0    \\
 2               & 31.7  & 19.0 & 4.3  & 1.7  & 0.9  & 0.6  & 0.1  \\
 4               & 38.0  & 36.0 & 20.7 & 5.2  & 2.3  & 1.3  & 0.4  \\
 6               & 40.3  & 39.7 & 36.9 & 21.3 & 5.6  & 2.6  & 0.4  \\
 8               & 41.4  & 41.2 & 40.3 & 37.3 & 21.6 & 5.8  & 0.7  \\
 10              & 42.1  & 42.0 & 41.6 & 40.6 & 37.6 & 21.8 & 0.9  \\
 20              & 43.6  & 43.5 & 43.5 & 43.4 & 43.3 & 43.1 & 22.1 \\
 \hline
\end{tabular}
\end{table}

\begin{table}
 \caption{The relative polarization degree
 $p_{rel}=p({\bf n},{\bf B})/p(\mu,0)$ as a function
 of parameters $a$ and $b$ at $C=1$.}
 \label{tab3}
 \begin{tabular}{@{}lccccccc}
 \hline
 $a\backslash b$ & 0     & 2     & 4     & 6     & 8     & 10    & 20 \\
 \hline
 0               & 0.5   & 0.354 & 0.224 & 0.158 & 0.121 & 0.098 & 0.050 \\
 2               & 0.354 & 0.334 & 0.236 & 0.165 & 0.124 & 0.100 & 0.050 \\
 4               & 0.224 & 0.236 & 0.246 & 0.186 & 0.135 & 0.106 & 0.051 \\
 6               & 0.158 & 0.165 & 0.186 & 0.203 & 0.158 & 0.118 & 0.052 \\
 8               & 0.121 & 0.124 & 0.135 & 0.158 & 0.176 & 0.140 & 0.054 \\
 10              & 0.098 & 0.100 & 0.106 & 0.118 & 0.140 & 0.158 & 0.057 \\
 20              & 0.050 & 0.050 & 0.051 & 0.052 & 0.054 & 0.057 & 0.112 \\
 \hline
\end{tabular}
\end{table}

\begin{table}
 \caption{The position angles $\chi^{\circ}$ as a function of
 parameters $a$ and $b$ at $C=1$.}
 \label{tab4}
 \begin{tabular}{@{}lccccccc}
 \hline
 $a\backslash b$ & 0     & 2    & 4    & 6    & 8    & 10   & 20 \\
 \hline
 0               & 0     & 0    & 0    & 0    & 0    & 0    & 0    \\
 2               & 22.5  & 15.9 & 6.6  & 3.1  & 1.8  & 1.1  & 0.3  \\
 4               & 31.7  & 29.1 & 19.0 & 8.4  & 4.3  & 2.6  & 0.6  \\
 6               & 35.8  & 34.8 & 30.9 & 20.1 & 9.2  & 4.8  & 0.9  \\
 8               & 38.0  & 37.6 & 36.0 & 31.7 & 20.7 & 9.7  & 1.3  \\
 10              & 39.3  & 39.1 & 38.4 & 36.6 & 32.2 & 21.1 & 1.9  \\
 20              & 42.1  & 42.1 & 42.0 & 42.0 & 41.7 & 41.2 & 21.8 \\
 \hline
\end{tabular}
\end{table}

The detailed numerical calculation (particularly presented in
tables 1 - 4) demonstrate that the existence of ${\bf B}_{\bot}$
can increase the relative degree of polarization $p_{rel}$
compared with the case $b=0$. This effect takes place for $a\ge
1-3$ and $b\le a$. The maximum polarization occurs at $b=a$ (see
tables 1 and 3). This effect stems some "resonant" regions in an
accretion disk where the Faraday rotation from ${\bf B}_{\|}$ is
balanced by opposite rotation from a perpendicular magnetic field
${\bf B}_{\bot}$. For $b>a$ polarization decreases quickly with
growing of parameter $b$. Note also that for $b=a\gg 1$ the
position angle $\chi \to 22.5^{\circ}$. For this case the
transition of $\chi$ from $\approx 45^{\circ}$ to very small value
occurs in the layer $|a-b|\approx 2-5$, i.e. we observe the
rotation of position angle $\chi$ from $\pm 45^{\circ}$ (depending
on the magnetic field direction) to $\chi \approx 0$.

Presented expressions and tables allow us to estimate the values
of polarization degree and position angle. We shall use these
expressions to explain the effect of the cosmological rotation of
the position angle of polarized QSOs discovered by
\citet{hutsemekers05}.

\section{Scenario of intrinsic mechanism}

The scenario of our explanation is the following. First of all, we
assume that at early time of the QSOs evolution, corresponding to
$z>2-3$, the magnetic field in accretion disk is directed along
the normal ${\bf N}$ to the surface, and the perpendicular
magnetic field ${\bf B}_{\bot}$ is practically absent. The
increase of ${\bf B}_{\bot}$ inside the accretion disk can be
considered as due to large-scale diffusion process, i.e. one can
assume ${\bf B}_{\bot}\sim Dt$. The time difference $t$ is
proportional to cosmological parameter $z$ (remember that large
$z$ correspond to early time of the evolution of the Universe).
The high values of ${\bf B}_{\|}$ ($a\ge 4$) correspond to
inclination angle $\chi \approx\pm 45^{\circ}$ of electric field
of the emerging radiation, as compared to the usual Thomson
position angle, which is parallel to accretion surface. The sign
plus corresponds to the case where ${\bf B}_{\|}$ is directed
outside the surface. The minus sign corresponds to opposite field
direction. As a result of large-scale diffusion magnetic field
${\bf B}_{\bot}$ acquires the values $B_{\bot}\approx B_{\|}$, and
the case $b\approx a >4 $ occurs. In small interval of growing of
parameter $b$ with $|a-b|\approx 2 - 5$ , where position angle of
outgoing radiation changes from initial value $\approx 45^{\circ}$
to final zero value, the angle of rotation can be roughly
approximated by linear dependence on value $b\sim t\sim z$.
Remember that usually position angle $\chi $ is observed with
fairly high error interval. So, evolution mechanism of the
increase of perpendicular magnetic field inside of accretion disk
presents simultaneously the intrinsic mechanism of rotation of
observed position angle, which is proportional to cosmological
parameter $z$. Note that this mechanism is restricted by
$45^{\circ}$ rotation.

The cosmological rotation of mean position angle, discovered by
\citet{hutsemekers05}, corresponds to approximately linear growing
from zero at $z=2$ to $\approx 90^{\circ}$ at $z\approx 0$. Our
mechanism can explain only $45^{\circ}$ rotation. Evidently there
is an additional mechanism of rotation, connected with the
alignment of quasar polarization vectors discovered also by
\citet{hutsemekers98}. Clearly, this alignment is due to some type
of anisotropy of QSOs distribution in cosmic space. For example,
we may assume that a part of quasars distribution presents spiral
type form. In this case the local position $\chi=0$ variates from
one quasar to another. The simultaneous action of our intrinsic
mechanism and variation of zero position $\chi=0$ for particular
quasars, corresponding to $z$-parameter, can explain the observed
effect of rotation of the mean position angle.

The mentioned maximum angle of rotation  $\chi \approx 45^{\circ}$
due to intrinsic mechanism may be not realized in reality. In this
case main part of observed total rotation will be due to kinematic
mechanism of variations of the normals to accretion disks surfaces
for particular quasars, with corresponding $z$ - parameters. We
only demonstrated  above that the intrinsic mechanism exists, and
it can explain rotation $\leq 45^{\circ}$. According to
\citet{hutsemekers05} the mean degree of polarization of all the
sample of quasars is equal to $1.38\%$. They mentioned also that
the alignment effect is more efficient for the low polarization
quasars than for the high polarization ones. This means that we
can consider the rotations of position angle of quasars with the
degree of polarization lesser then the value $1.38\%$.

Now we give some examples of values of rotation for $i=45^{\circ},
60^{\circ}$ and $80^{\circ}$, using our tables 1 and 2, and that,
according to \citet{chandrasekhar50}, the polarization degrees
$p(\mu,0)$ for these inclination angles are 1.1\%, 4.04\%, and
6\%, respectively. The accretion disk in the initial state
($a=4,b=0$) has degrees of polarization 0.27\%, 0.98\% and 1.45\%,
respectively, and the position angle $\chi=38^{\circ}$. When the
growing perpendicular magnetic field $B_{\bot}$ gives rise to
$b=2$, accretion disk acquires the degrees of polarization 0.3\%,
1.09\%, and 1.63\% with position angle $\chi=36^{\circ}$, i.e we
have the rotation $\Delta \chi= 2^{\circ}$. For $b=4$ the
corresponding values are 0.39\%, 1.42\%, 2,1\% with position angle
$\chi=20.7^{\circ}$, i.e. have rotation $\Delta\chi =17.3^{\circ}$
from the state ($a=4, b=0$), and the rotation $\Delta\chi
=15.3^{\circ}$ from the state ($a=4, b=2$). Remember that values
of $b$ depend on $z$-parameter, and belong to quasars with various
$z$, which are at different distances from an observer. Remind
also that optical radiation escapes from accretion surface far
from the centre of disk and magnetic field in this place is much
lesser than that in the centre. The estimations give rise to the
values $B\approx 10^3 - 10^4$ G in the central parts of a disk.
The used value $a=4$ correspond to $B_{\|}\simeq 16.5\mu$~G for
$\lambda =0.55\mu$m. It seems the process of diffusion of magnetic
field from central part to the places where optical radiation
arises can be slow. The increase of polarization degree with the
grow of parameter $b$ (up to $b=a$) is new effect which was
explained in the end of the previous section.

Thus, the observed effect of the cosmological rotation of
polarization vectors of QSOs can be explained partly by evolution
of a magnetic field in accretion disks. As a result of such
evolution the topology of magnetic field in AGNs is changed and
implies the transition, for example, from the predominant vertical
domain distribution to the predominant horizontal domain
distribution. The ratio of domain sizes can be changed respect to
the cosmological redshift $z$.

The physical mechanism of magnetic inversion considered here can
be the same one which was considered recently by
\citet{igumentsev09} for explaining the spectral transition of
black holes binaries. The base for these phenomena can be the
development of a magnetically arrested accretion disk attributable
to the accumulation of a vertical magnetic field in a central part
of this disk. The development and evolution of powerful jets
provides also the evolution and inversion of magnetic field
components in an accretion disk.

More interesting mechanism has been recently suggested by
\citet{lyutikov09}. Strong magnetic field can modify motion in the
curved space-time of spinning black holes and change the stability
conditions of circular orbits. Magnetocentrifugal jet launching
from accretion disks around black holes is also connected to
topology of accretion disk magnetic field. According to
\citet{lyutikov09}, magnetocentrifugal launching for a
Schwarzschild black hole requires that the poloidal component of
magnetic field makes an angle less that $60^{\circ}$ to the
outward direction at the disk surface. For the prograde rotating
disks around Kerr black holes this angle increases and becomes
$90^{\circ}$ for footpoints anchored to the inner region of an
accretion disk for a limit spinning $a_* = 1$ black hole ($a_* = a
/ M_{BH}$). It means that the effect of cosmological rotation of
QSO polarization plane can be interpreted as a result of spin
evolution of QSO: $a_* \approx 1$ for $z > 1$ and $a_* \ll 1$ for
$z \ll 1$. This suggestion has some evidence. Recently, the
constraints on the spins of the black holes in the nearby QSOs
have been obtained: $a_* = 0.6 \pm 0.2$ in the narrow-line Seyfert
SWIPT 12127+5654 \citep{miniutti09}, $a_* = 0.60 \pm 0.07$ for the
SMBH in Fairall 9 \citep{schmoll09}.

The another idea has been developed by \citet{contopoulos09}. They
developed the scenario in which cosmic magnetic fields are
generated near the inner edges of accretion disks in AGNs by
azimuthal electric currents due to difference between the plasma
electron and ion velocities that arises when the electrons are
retarded by interactions with X-ray photons. This mechanism namely
relates the polarity of poloidal magnetic field to the angular
velocity of the accretion disk, i.e. this polarity depends
strongly on spin of a black hole. If the spin decreases at low
redshifts $z < 1$, this effect produces magnetic inversion in the
accretion disks.

Unfortunately, if the effect of rotation of polarization planes in
QSOs can be partly explained in framework of our mechanism, the
origin of cosmological alignment of polarization vectors becomes
open. It is not except, that the mechanism of cosmological
differential rotation produces such kind of alignment. It seems
more probable supposition of alignment mechanism is the existence
of turbulent motions of the huge-scale.

Recently, the interesting idea of alignment of galaxies has been
suggested by \citet{trujillo07}. They claimed that galaxies are
not distributed randomly throughout space but are arranged in
''cosmic web'' of filaments and walls. Unfortunately, there is
still no compelling observational evidence of a link between the
structure of the cosmic web and how galaxies form within it. The
basis for this connection is the origin of galaxy angular
momentum. The spin of galaxies can be generated by tidal torques
operating in the early Universe on primordial material from which
galaxy is formed. In its order, magnetic fields in QSOs and
galaxies can correlate with galaxy rotation rate and its angular
momentum. \citet{trujillo07} claimed that observational link
between large-scale structure and the properties of individual
galaxies has been definitively established. One should remind the
result by \citet{borquet08}. They found a correlation between the
polarization position angle and the position angle of the major
axis of the host galaxy extended emission. It seems that in the
frame of this idea our intrinsic mechanism can explain the total
rotation of the mean position angle if we shall suppose that
position of zero's $\chi$ in neighboring "webs" has a jump
$\approx 45^{\circ}$.

\section{Conclusions}

We showed that the observed by \citet{hutsemekers05} effect of
cosmological rotation of polarization vectors of QSOs can be
partly explained by the evolution of magnetic field in accretion
disks around SMBHs. The evolution of topology of magnetic field in
AGNs produces the transition from the predominant vertical domain
distribution at $z \geq 1$ to the predominant horizontal domain
distribution at $z < 1$. Such kind of transition can be explained
by the cosmological evolution of spin of supermassive black holes
from the Kerr case at $z \geq 1$ to Schwarzschild case at $z < 1$.
The calculations show that this effect is able to provide the
observed rotation measure at the level of $30^{\circ}$ / Gpc. The
alignment effect itself is leaving unexplained. One of possible
scenarios is alignment of spins of galaxies with observed
large-scale structure of the Universe, which possibly represent
the turbulent eddies of the huge scale.

\section*{Acknowledgements}

This research was supported by the RFBR (project No.
07-02-00535a), the program of Prezidium of RAS ''Origin and
Evolution of Stars and Galaxies'', the program of the Department
of Physical Sciences of RAS ''Extended Objects in the Universe''
and by the Grant from President of the Russian Federation ''The
Basic Scientific Schools'' NS-6110.2008.2. M.Yu. Piotrovich
acknowledges the Council of Grants of the President of the Russian
Federation for Young Scientists, grant No. 4101.2008.2.

\end{document}